\documentclass[twocolumn]{aastex631}

\newcommand{\kpc}{\ensuremath{\mathrm{kpc}}}
\newcommand{\kms}{\ensuremath{\mathrm{km}~\mathrm{s}^{-1}}}
\newcommand{\kpckms}{\kpc~\kms}
\newcommand{\kpckmsR}{\ensuremath{(\kpc~\kms)^{1/2}}}
\newcommand{\Msun}{\ensuremath{\mathrm{M}_\odot}}

\shorttitle{Snails in Simulations}
\shortauthors{Darragh-Ford et al.}
\graphicspath{{./}{figures/}}

\begin{document}

\title{\texttt{ESCARGOT}: Mapping Vertical Phase Spiral Characteristics Throughout the Real and Simulated Milky Way}

\author[0000-0002-8800-5652]{Elise Darragh-Ford}
\affiliation{Kavli Institute for Particle Astrophysics and Cosmology and Department of Physics, Stanford University, Stanford, CA 94305, USA}
\affiliation{SLAC National Accelerator Laboratory, Menlo Park, CA 94025, USA}

\author[0000-0001-8917-1532]{Jason A. S. Hunt}
\affiliation{Center for Computational Astrophysics, Flatiron Institute, 162 5th Av., New York City, NY 10010, USA}

\author[0000-0003-0872-7098]{Adrian M. Price-Whelan}
\affiliation{Center for Computational Astrophysics, Flatiron Institute, 162 5th Av., New York City, NY 10010, USA}

\author[0000-0001-6244-6727]{Kathryn V. Johnston}
\affiliation{Department of Astronomy, Columbia University, New York, NY 10027, USA}


\begin{abstract}
The recent discovery of a spiral pattern in the vertical kinematic structure in the solar neighborhood provides a prime opportunity to study non-equilibrium dynamics in the Milky Way from local stellar kinematics. Furthermore, results from simulations indicate that even in a limited volume, differences in stellar orbital histories allow us to trace variations in the initial perturbation across large regions of the disk. We present \texttt{ESCARGOT}, a novel algorithm for studying these variations in both simulated and observed data sets. \texttt{ESCARGOT} automatically extracts key quantities from the structure of a given phase spiral, including the time since perturbation and the perturbation mode. We test \texttt{ESCARGOT} on simulated data and show that it is capable of accurately recovering information about the time since the perturbation occurred as well as subtle differences in phase spiral morphology due to stellar location in the disk at the time of perturbation. We apply \texttt{ESCARGOT} to kinematic data from data release 3 of the \emph{Gaia} mission in bins of guiding radius. We show that similar structural differences in morphology occur in the \emph{Gaia} phase spirals as a function of stellar orbital history. These results indicate that the phase spirals are the product of a complex dynamical response in the disk with large-scale coupling between different regions of phase space.  
\end{abstract}

\keywords{\href{http://astrothesaurus.org/uat/589}{Galaxy disks (589)}; \href{http://astrothesaurus.org/uat/1509}{Solar neighborhood (1509); \href{http://astrothesaurus.org/uat/1050}{Milky Way disk (1050)}; \href{http://astrothesaurus.org/uat/602}{Galaxy kinematics (602)}; \href{http://astrothesaurus.org/uat/594}{Galaxy evolution (594)}}}


\section{Introduction} 
\label{sec:intro}
One of the most exciting results from data release 2 (DR2) of the \emph{Gaia} Mission \citep{DR2} was the discovery of a one-armed spiral pattern (hereafter, a ``phase spiral'') in vertical kinematic ($z,v_z$) phase space through the projection of the 6-dimensional coordinates for stars near the Sun \citep{Antoja_2018}. 
This clear signature of disequilibrium in the local stellar kinematics provides the exciting demonstration that the Galactic disk can be used to study the large-scale drivers of non-equilibrium dynamics in the Milky Way \citep[as suggested by, e.g.,][]{Widrow:2012}. The phase spiral is formed from the phase mixing of a disk population that has been offset perpendicular to the Galactic plane from a random distribution at some point in the past. The systematic increase in vertical time periods for orbits that explore further from the disk plane means that their offset becomes progressively lagged relative to orbits closer to the disk plane over time. In this simple picture, the morphology of the phase spiral contains information about the time since the perturbation occurred \citep[as first calculated by][]{Antoja_2018} as well as the frequency ratios, and hence mass distribution perpendicular to the disk \citep[e.g.][]{widmark_21}. The picture is complicated by the self-gravity of phase-mixing population itself (i.e. the disk), whose vertical oscillations can drive further phase spirals to form \citep{darling_19,Bland_Hawthorn_2021}.

While the exact origins of such phase spirals is an ongoing topic of debate, simulations have produced qualitatively similar structures through global interactions between the Milky Way and a dwarf galaxy with mass $\sim 10^{10}~\Msun$ \citep{Laporte_2019,Bland_Hawthorn_2021, Hunt_2021}. These results point to the Sagittarius dwarf galaxy as a plausible driver of this signature. However this picture is complicated by present day estimates of Sagittarius' mass, which suggest a light remnant \citep[$\sim4\times10^8~\Msun$;][]{Vasiliev_20}, at odds with the degree of vertical disequilibrium observed in the Milky Way, and by the fact that attempts to reproduce local vertical asymmetry with Sagittarius alone have been unsuccessful \citep{bennett_22}.

Regardless, any weak perturbation about the (local) midplane of the mass distribution in the Galaxy that distorts the population away from equilibrium will phase-mix away in a similar manner, naturally leading to one-armed (if the response is asymmetric) or two-armed (if the response is symmetric) phase spirals \citep{Hunt_2021,Banik_2022}.  
Other plausible origins discussed so far include the buckling of a galactic bar \citep{khoperskov_19},  the creation of a large-scale dark matter wake \citep{Grand2022}, or noise in the disk from giant molecular clouds \citep{tremaine_23}.

Luckily, there exist multiple phase spirals to help distinguish between these scenarios. When samples local to the Sun are divided into different groups according to orbital properties (e.g. $z$-angular momentum or azimuthal action, $J_\phi$, and azimuthal angle,\footnote{I.e. the conjugate angle to the azimuthal action, not the present-day azimuth $\phi$.} $\theta_\phi$) there are clear differences in the morphologies of the $z-v_z$ phase spiral between them \citep{Li_2020,Hunt_2021,Gandhi_2022}, including the transition from one- to two-armed morphologies in the lower angular momentum orbits that explore regions closer to the Galactic center \citep{Hunt_2022}. 
These differences between orbit groups can also be attributed to phase mixing following a perturbation, this time in location across the Galaxy. Groups that are currently close in Galactocentric radius and azimuth $R,\phi$ in the Galactic disk were not always close. Hence, they may have experienced a given perturbation at a different time or with a different force history. Or, in some cases, they may have been perturbed by an entirely different event \citep[see][for a longer discussion]{Gandhi_2022}.

With the increasing scope of observational data -- in number, accuracy, and volume explored -- from the \emph{Gaia} mission data release 3 (DR3) \citep{DR3} \citep[and future data from Sloan Digital Sky Survey V--Milky Way Mapper][]{SDSSV} comes the opportunity to make detailed maps of the morphology of phase spirals as a function of orbital properties --- both action and phase. Conceptually, such a map might be the key to understanding the complex perturbation history of the Milky Way from internal and external perturbers. A perturbation arising from the ends of a buckling bar that are diametrically opposed across the inner Galaxy should give rise to a pattern of correlated phase spirals across the disk that is distinct from those originating from a local perturbation in the Galactic outskirts due to an infalling satellite.
However, in order to fully take advantage of the information offered by these signatures to reconstruct a nuanced dynamical history of the Milky Way, we need a robust method for accurately characterizing these phase spirals as a function of location in the Galaxy. 

In this work, we introduce \texttt{ESCARGOT} (Estimating Spiral Characteristics and Recovering General Orbital Timescales) an algorithm designed to automatically extract and characterize phase spirals. \texttt{ESCARGOT} shares some conceptual similarities with the algorithms of \cite{antoja_22} and \cite{frankel_22} which were developed concurrently and independently, although specific choices in implementation differ. In Section \ref{sec:spiral_charac} we describe the characteristics of the phase spirals and in Section \ref{sec:algorithm} we describe the construction of the \texttt{ESCARGOT} algorithm to extract them. In Section \ref{sec:simulation_results} we validate \texttt{ESCARGOT}'s performance on simulated data before showing its results on the recently released \emph{Gaia} DR3 catalogs in Section \ref{sec:gaia_data}. In Section \ref{sec:discussion} we discuss our findings compared to the literature, and in Section \ref{sec:conclusion} we state our conclusions.

\begin{figure*}
\includegraphics[width=\textwidth]{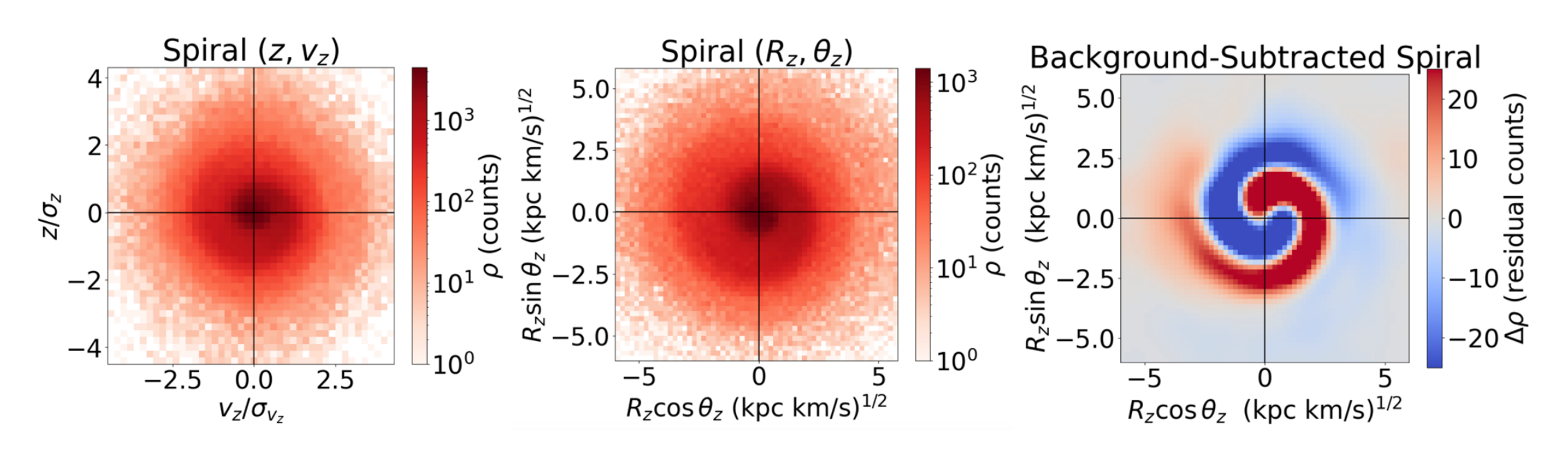}
\caption{The \emph{Gaia} phase spiral in stellar number counts (left and center panels) and a Gaussian-smoothed and background-subtracted representation (right panel). The coordinates in the middle panel are computed by transforming to action-angle coordinates $J_z, \theta_z$ and defining $R_z \equiv \sqrt{J_z}$. The color bar in the right panel has units of residual stellar number counts (and is linearly scaled).} 
\label{fig:density_plots}
\end{figure*}

\begin{figure*}
\centering
\includegraphics[width=0.75\textwidth]{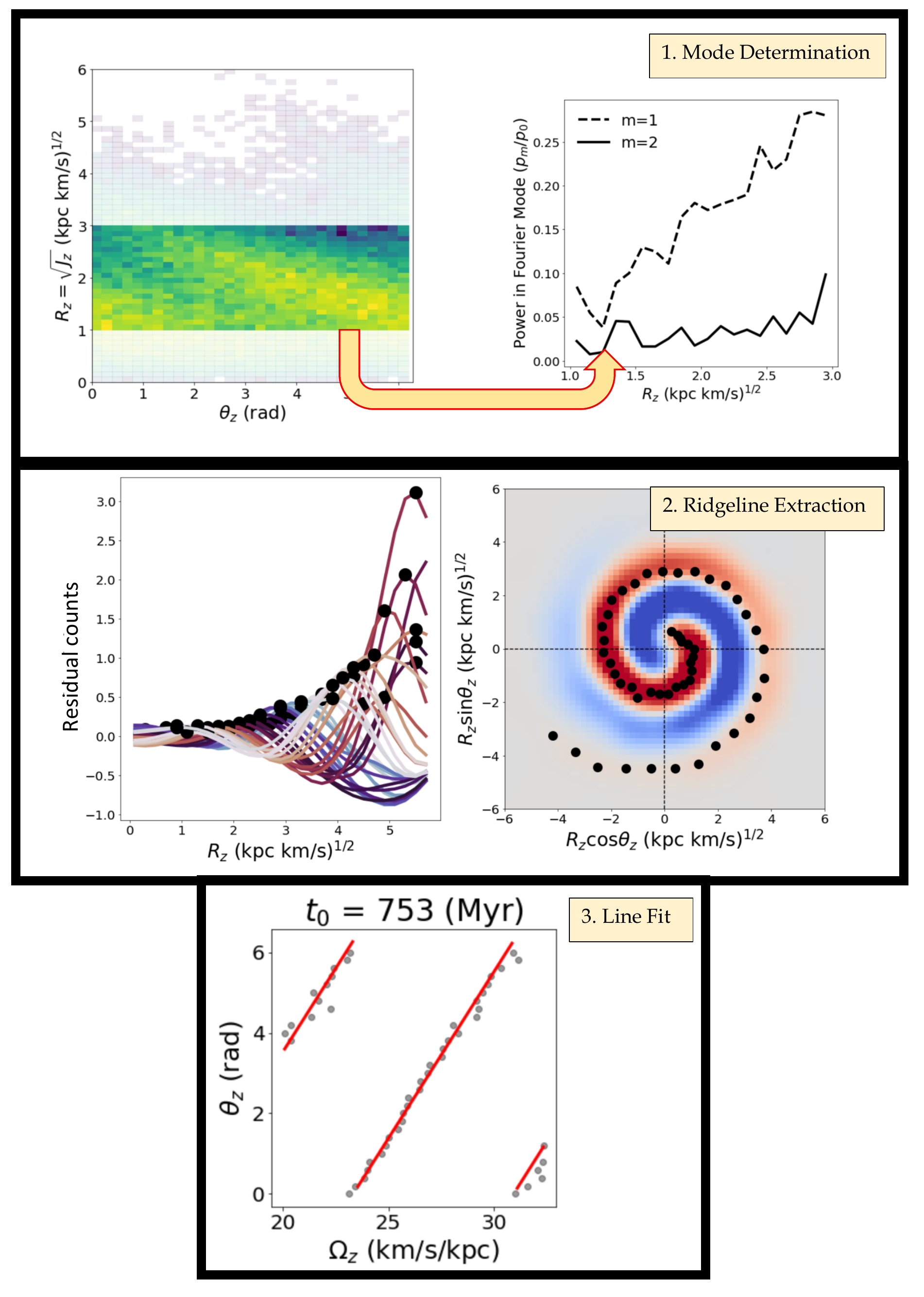}
\caption{Steps in \texttt{ESCARGOT}. \emph{top:} mode determination using a Fourier transform. Stars are binned in $R_z$--$\theta_z$ (\emph{left}). For each bin in $R_z$, we compute a Fourier transform with respect to $\theta_z$ and measure the strength of the $m=1$ and $m=2$ modes relative to the background $m=0$ mode (\emph{right}). \emph{middle:} ridgeline extraction using a peak finding algorithm. The peaks are extracted in wedges of $\theta_z$ (colored lines, \emph{left}) with a minimum threshold of 0.05 in residual counts. The peaks are plotted as black dots as a function of $R_z$ (\emph{left}) and on top of the background subtracted phase spiral (\emph{right}). \emph{bottom:} a fit to the ridgeline using the Hough transform. We find a best-fit line (red) to the extracted ridgeline (blue points) using a combination of the Hough transform and least-square error.}
\label{fig:algorithm}
\end{figure*}

\section{Phase Spiral Properties and Measurables}
\label{sec:spiral_charac}
We characterize phase spirals in action-angle ($J_z,\theta_z$) rather than physical ($z,v_z$) coordinates where they were first discovered. Action-angle coordinates are canonical coordinates in which the momenta ($J$) are integrals of motion, meaning they are constant along a given orbit and constrain the shape of the orbit. Orbits in action-angle space have a constant frequency ($\Omega$) and are fully characterized by their actions ($J$) and angles ($\theta$). The downside of working in action-angle space is that we need to assume a gravitational potential in order to calculate actions and angles. In addition, actions are only well-defined in the limit of an adiabatically changing potential, and the transformation to action-angle space can be ill-defined in the presence of large perturbations. However, previous work has shown that computing actions for stars in \emph{Gaia} DR2 using an approximate Milky Way potential, where we implicitly assume the measured phase-structure is only a weak perturbation to a well-defined background potential, shows well-resolved $J_z,\theta_z$ phase spirals \citep{Bland_Hawthorn_2019}. 

We choose to focus on action-angle coordinates in this work because the winding of the $z,v_z$ phase spiral in these coordinates is given by the simple equation 
\begin{equation}
\label{eq:eq1}
    \theta_z = \theta_0 + \Omega_z \,t_0
\end{equation}
where $\theta_z, \Omega_z$ are the $z$-angle and vertical frequency components respectively, $\theta_0$ is the offset angle of the initial perturbation and $t_0$ is the time since perturbation (e.g., time since the satellite encounter) \citep{Binney_2018}. This allows us to characterize the phase spiral using three numbers:
\begin{enumerate}
    \item $t_0$: the time since interaction
    \item $\theta_0$: the initial pitch angle of the perturbation
    \item $m$: the mode of the phase spiral
\end{enumerate}
where $t_0$ and $\theta_0$ come directly from Equation \ref{eq:eq1} and $m$ describes the dominant mode of the phase spiral. This can either be an $m=1$ one-armed spiral, or an $m=2$ two-armed spiral. For phase spirals introduced by an interaction with an external perturber, previous work has shown that the mode of the induced phase spirals depends on the speed of the satellite interaction. Faster interactions tend to excite ``breathing'' modes and two-armed phase spirals, whereas slower interactions tend to produce ``bending'' modes and one-armed phase spirals \citep{widrow_14,Hunt_2021, Banik_2022}. Both $m=1$ and $m=2$ phase spirals have been found within the Milky Way \citep{Hunt_2022}.

We visualize the phase spiral in action-angle coordinates using the intermediate coordinates $R_z\,\cos{\theta_z}, \,R_z \, \sin{\theta_z}$  where $R_z = \sqrt{J_z}$ and has units of $(\kpckms)^{1/2}$). The area in this plane is roughly equivalent to integrating in the $z,v_z$ plane. In both cases, this area has units of action \citep{Binney_2018}. An example of the phase spiral density in $z,v_z$ and $R_z,\theta_z$ can be seen in Figure~\ref{fig:density_plots} (left and center panels, respectively).  Here, we normalize $z$ and $v_z$ by the standard-deviation to make the phase spiral look more circular. However, it is clear by eye that the phase spiral in the left panel is still oblate and shows an $m=4$-like distortion in the outer regions.  The phase spiral in the center, which is plotted using action-angle coordinates, is more circular.

The background subtracted image is created by randomizing the $\theta_z$ values for the stars in the center panel, and subtracting the randomized density distribution from the unrandomized distribution. For visual clarity, the background subtracted distribution is then smoothed with a 2D Gaussian kernel ($\sigma = 2.5$ pixels). The background subtraction shows the spiral pattern as an over(under)density without the smooth density background. 

\section{Introducing \texttt{ESCARGOT}} 
\label{sec:algorithm}

The morphological information for a given phase spiral is extracted with a multi-step algorithm. The main steps of \texttt{ESCARGOT} are summarized as follows: 

\begin{enumerate}
    \item Extract the dominant mode ($m$) using a Fourier Transform
    \item Perform background subtraction and ridgeline extraction using a peak-finding algorithm
    \item Perform a linear fit $(t_0, \theta_0)$ to the ridgeline using the Hough transform 
    
\end{enumerate}

Each of these steps is detailed more thoroughly in the following sections along with the specific parameter choices we have made. A graphic representation of each of the steps of \texttt{ESCARGOT} can be seen in Figure~\ref{fig:algorithm}.

\subsection{Mode Extraction}
To extract the dominant mode we generate a 2D histogram as a function $\theta_z$ and $R_z$, as shown in the top left panel of Figure \ref{fig:algorithm}. We construct 21 bins in $R_z$ between $R_{\rm min} < R_z < R_{\rm max}$. Where $R_{\rm min} = 1.0~\kpckmsR$ and $R_{\rm max} = 3.0~\kpckmsR$.  We focus on the middle portion (in $R_z$) of the phase spiral as it is seen consistently to have the cleanest signal. $\theta_z$ values are binned with a bin width of 0.2 rad. 

We then apply a Gaussian smoothing kernel ($\sigma = 1.0$ pixels) and compute the Fourier Transform for each of the $R_z$ bins and sum over the $m=1$ and $m=2$ modes for each bin in $R_z$, normalized by the strength of the $m=0$ mode in that bin:
\begin{equation}
    P_m = \sum_{i=R_{\rm min}}^{R_{\rm max}} \frac{p_i (m=m)}{p_i (m=0)}) 
\end{equation}
where $p_i (m=m)$ is the power in the $m=m$ mode of the Fourier transform for the $i^{\rm th}$ bin in $R_z$. 

The dominant spiral mode is determined as $\mathrm{max}(P_1, P_2)$. The power in the normalized $m=1$ and $m=2$ modes as a function of $R_z$ can be seen in the top right panel of Figure \ref{fig:algorithm}. Figure \ref{fig:algorithm} shows an $m=1$ phase spiral, the $m=1$ mode dominates at all radii.

\subsection{Ridgeline Extraction}

We next identify the ridgeline of peak density along the spiral. We generate a 2D histogram using the $R_z,\theta_z$ values, this time with a bin width of 0.2 [$(\kpckms)^{1/2}$, rad] in both dimensions out to $R_z < 6.0$ $(\kpckms)^{1/2}$. The number counts in each bin $i,j$ of this histogram are $N_{ij}$.  We also construct a second histogram in the same manner, but after randomly shuffling the $\theta_z$ values ($N_{ij, {\mathrm{rand}}}$).

We use $N_{ij,{\mathrm{rand}}}$ as an estimate of the background density. We compute the mean background density $(\mu_{i, {\mathrm{rand}}})$ as a function of $R_z$ by averaging over the $\theta_z$ bins and then subtract it off: 
\begin{equation}
    \Delta N_{ij} = N_{ij} - \mu_{i, {\mathrm{rand}}}
\end{equation}
We then smooth the residual density field ($\Delta N_{ij}$) with a Gaussian kernel ($\sigma = 2.5$ pixels). An example of a background-subtracted and smoothed phase spiral can be seen in the right panel of Figure~\ref{fig:density_plots}, here plotted in $R_z \, \cos{\theta_z}, R_z \, \sin{\theta_z}$ rather than $R_z$, $\theta_z$ to show the spiral structure. 

Once we have the residual density field, we extract the ridgeline using a peak-finding algorithm with a threshold of 0.05 residual counts. We only consider peaks with $0.5 < R_z~(\kpckms)^{1/2} < 5.5$ to exclude poorly resolved spiral structure at the center or outskirts of the phase spiral. The results of the peak-finding algorithm can be seen on the \emph{left} side of Figure~\ref{fig:algorithm} where the black dots mark the extracted peaks. The extracted peaks are then plotted over the residual density spiral on the \emph{right} of Figure~\ref{fig:algorithm}. As can be seen by eye, the peaks accurately trace the ridgeline of the phase spiral. 

\subsection{Ridgeline Fit}
Having extracted the ridgeline in $\theta_z-R_z$, we then convert $R_z$ to $\Omega_z$ by selecting the mean value of $\Omega_z$ within $R_z \pm 0.1$ $(\kpckms)^{1/2}$. We then fit a line to $\theta_z,\Omega_z$ values of the extracted ridgeline points using the Hough transform where the line is constrained to have a positive slope.  Out of the top 10$^{\rm th}$ percentile of fits returned by the Hough transform, the best fit is selected using a simple least-square error from the points of the extracted ridgeline. The reason for this two step process is that discontinuities due to wrapping in $\theta_z$ cause traditional line fitting methods to struggle to converge to reasonable parameter values. However, the Hough transform still only fits to the dominant ridgeline, ignoring secondary wrappings in $\theta_z$. These secondary wrappings will be included in the error calculation using the least-square error. We also apply a regularization term to the least-square calculation to avoid overfitting, which as calculated as $0.5 \times p_0^2$ where $p_0$ is the slope of the line (equivalent to assuming a Gaussian prior on the slope, centered on zero, but with unit variance). The result of the line-fitting procedure can be seen in the bottom panel of Figure~\ref{fig:algorithm} where the blue points are the extracted ridgeline in $\Omega_z, \theta_z$ and the red line is our best fit. We estimate the error on this fit by bootstrap resampling the ridgeline points, where the number of resampled points is the same as the number of ridgeline points we extract, and refitting the line, the error is calculated as the standard deviation of 20 realizations of boostrap resampling.

The full \texttt{ESCARGOT} algorithm returns two key numbers that quantify the shape of the phase spiral: (1) $m$, the dominant mode, which indicates whether the spiral is a one-armed or two-armed spiral and (2) $t_0$, the interaction time, which estimates the time since the perturbation occurred. For the phase spiral in Figure~\ref{fig:algorithm} the recovered mode is $m=1$ and $t_0 = 753 \pm 18$ Myrs. 

\begin{figure*}
\includegraphics[width=0.85\columnwidth]{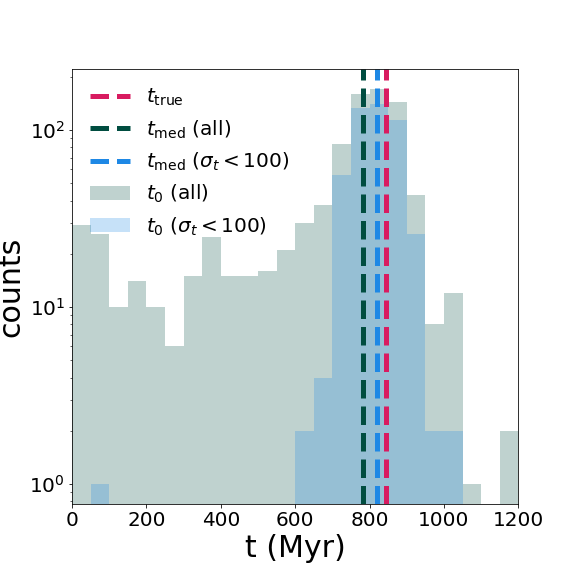}
\includegraphics[width=1.15\columnwidth]{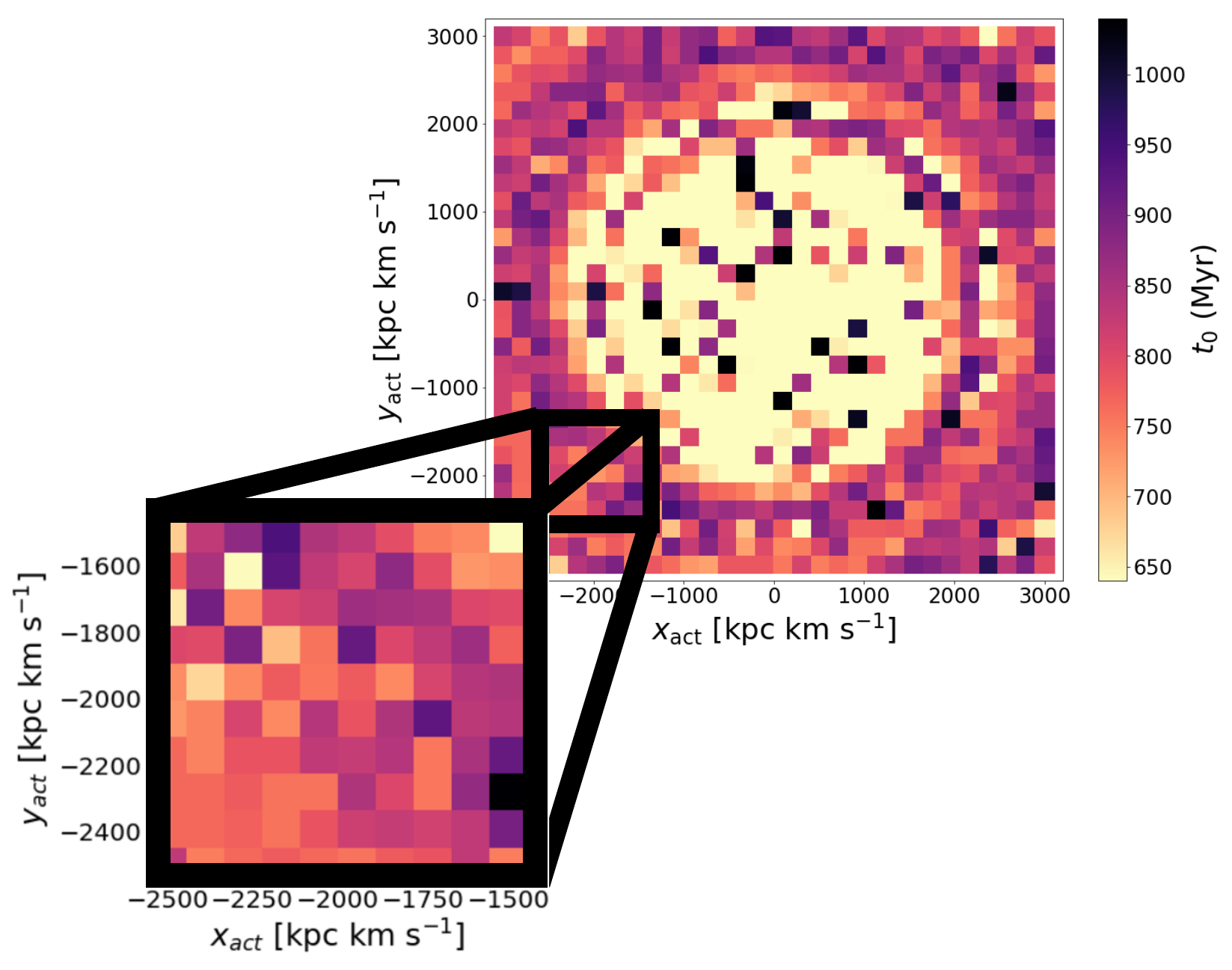}
\caption{\emph{left:} Distribution of interaction times, $t_0$, recovered from application of \texttt{ESCARGOT} to phase spirals in the idealized simulation. The dark green histogram shows all of the bins, while the blue shows only bins with a well-measured fit ($\sigma_t < 100$ Myr), which approximately corresponds to $2000 \lesssim J_{\phi}$ $\lesssim 4000$ (km s$^{-1}$ kpc$^{-1}$). The dashed lines mark the median of both the full (dark green) and restricted (blue) samples. The magenta dashed line shows the actual time since dwarf disk crossing. \emph{right:} Distribution of interaction times ($t_0$) recovered from the simulation as a function of $x_{act}$--$y_{act}$ (See equation \ref{eqn:act}.). The cutout displays a zoom-in on $-2500 < x_{\rm act}, y_{\rm act} < -1500$ (kpc km s$^{-1}$) to show the fine-grained structure.} 
\label{fig:times}
\end{figure*}

\begin{figure*}
\centering
\includegraphics[width=\textwidth]{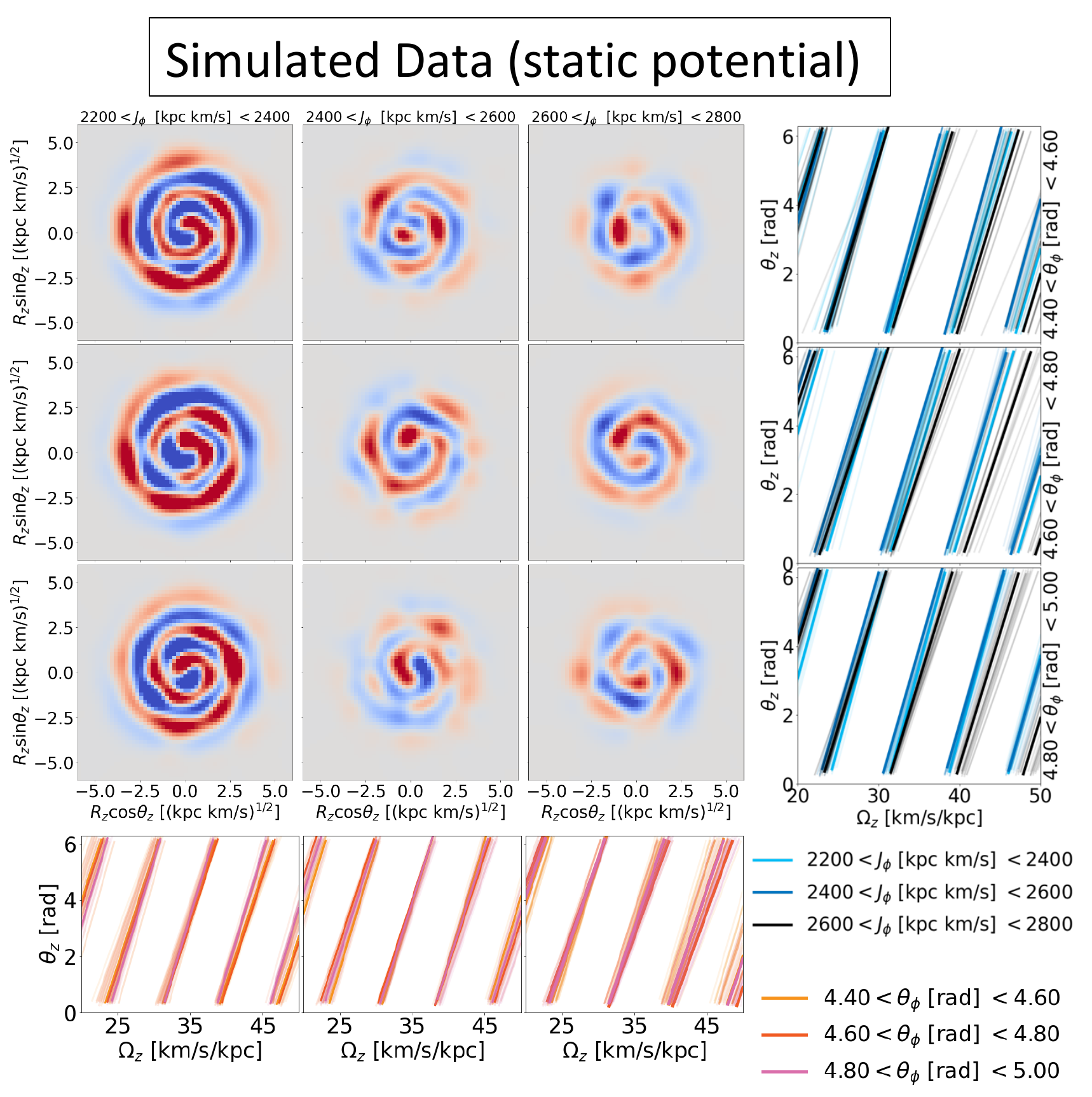}
\caption{\emph{Analysis of simulated data (static potential) --- center panel:} background subtracted phase spirals as a function of $J_\phi,-\theta_\phi$. \emph{frame:} Ridgeline fits to the spirals projected into $\theta_z - \Omega_z$ for different $\theta_\phi$ (right  panels) and $J_\phi$ (bottom panels) bins. The lighter lines represent the fit to the data after bootstrap resampling. The center panel shows the smooth variation in phase spiral morphology as a function of $J_\phi,-\theta_\phi$. In the bottom (left) panels the subtle shifts in phase spiral slope and pitch angle as a function of $\theta_\phi$ ($J_\phi$) can be seen.}
\label{fig:fits}
\end{figure*}

\begin{figure}
\includegraphics[width=\columnwidth]{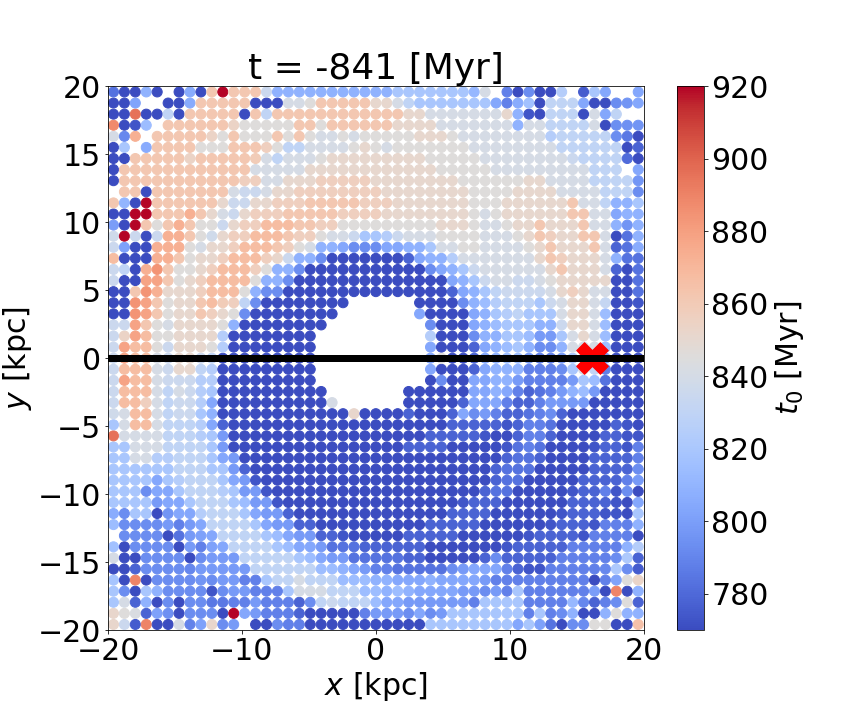}
\caption{Plot of disk at time of satellite crossing colored by estimated time since interaction at $z = 0$ ($t_0$) using only well-resolved phase spirals ($\sigma_t < 100$ Myr). The colorbar represents the estimated time since interaction at $z = 0$ with \emph{white} corresponding to $t_0 = t_{\mathrm{true}} = 844$ Myrs, \emph{red} corresponding to longer than expected interaction times and \emph{blue} corresponding to shorter than expected interaction times. The red X marks the disk-crossing of the satellite (the position at this snapshot), while the black line shows the orbital track in the X-Y plane. The disk is rotating counterclockwise in this projection.} 
\label{fig:traced_back}
\end{figure}

\section{Results I: Simulations}
\label{sec:simulation_results}
We test \texttt{ESCARGOT} using an idealized simulation of a Milky Way--dwarf interaction to demonstrate the performance of the algorithm. The simulation consists of a satellite galaxy that makes a single orbital passage through a thin disk of $10^9$ test particles in a static Milky Way potential model. 

The initial conditions for the test particle model are sampled from the \sc{quasiisothermaldf }\rm distribution function in \texttt{galpy}~ \citep{bovy_15}, adapted from \cite{binney_10} which is expressed in terms of action-angle variables. The distribution function is initialized with a radial scale length of 2.66 kpc, a local radial and vertical velocity dispersion of 110 and 16.5 km s$^{-1}$, respectively, and a radial scale length of 8 kpc for the velocity dispersions. The potential is evaluated with the \texttt{MWPotential2014} potential model in \texttt{galpy}. We generate $10^9$ position and velocity samples between radii of 1 and 13 kpc.

The dwarf galaxy is modeled using \texttt{galpy}'s \texttt{PlummerPotential}. It has a mass of $3\times10^9$ $M_\odot$, and a scale parameter of 1. The satellite present day position and velocity is set as $(x, y, z)=(25,0,300)$ kpc, $(v_x, v_y, v_z)=(0,0,-300)$ km s$^{-1}$. The orbit is integrated backwards in \texttt{MWPotential2014 } and then the dwarf potential is set to follow this orbit using the \texttt{MovingObjectPotential}.
 
 The simulation is evolved for 1.25 Gyrs with a disk crossing time occurring at $t_{\rm cross}=-844~\mathrm{Myr}$ (where $t=0$ is ``present day''). This disk crossing occurred at $R_{\rm cross} = 16$ kpc, corresponding to $J_{\phi, \rm cross}= 3245$ kpc km s$^{-1}$ for a circular orbit at that radius.  Due to the speed and orientation of the interaction only $m=1$ phase spirals are generated in the disk. 

This model is not meant to represent the Milky Way and Sagittarius, but is instead a simple idealized test case to see if we can correctly recover the impact time in a system that has undergone a single perturbation. Since every phase spiral originates from the same impact, there will be no effect from interaction with the static dark matter halo, no perturbation from dark matter subhalos or giant molecular clouds, and no self-gravity in the disc to affect the response.

We bin the stars in the disk by ``guiding center Cartesian coordinates", 
\begin{equation}
\label{eqn:act}
(x_{\mathrm{act}}, y_{\mathrm{act}}) = (J_\phi \, \cos{\theta_\phi}, J_\phi \, \sin{\theta_\phi})
\end{equation}
following \citep{Hunt_2020, Hunt_2021}, where $\theta_{\phi}$ is the conjugate phase angle to $J_{\phi}$. This transformation moves stars back to their guiding centers instead of their physical location, grouping stars with shared orbital histories. It is no longer a physical map of the disc, but instead a map of stars with similar orbits. This transformation is described in more detail in Section~\ref{sec:framework}, and \citep{Hunt_2020}. The transformation has also been shown to produce stronger phase spirals because the grouping by angular momentum, $J_{\phi}$ and azimuthal phase angle $\theta_{\phi}$, selects stars that were at similar positions in the disk at the time of interaction \citep{Hunt_2021,Gandhi_2022}. 

We run \texttt{ESCARGOT} on all stars selected in thirty bins of $200$ kpc km s$^{-1}$ in $x_{\mathrm{act}}$ and $y_{\mathrm{act}}$ between $[-3000, 3000]$ kpc km s$^{-1}$. The distribution of times recovered from the full simulated disk can be seen in the left panel of Figure \ref{fig:times}, where we also select a subset of fits where the standard deviation on the slope from bootstrap resampling is low ($\sigma_t < 100$ Myr). These well-resolved phase spirals correspond to a ring ($2000 \lesssim J_{\phi}$ $\lesssim 4000$ km s$^{-1}$ kpc$^{-1}$) in the the outer reaches of the disk where visual examination shows stronger phase spirals. This ring is also $\sim J_{\phi, \rm cross}$. While, the median of both samples (788 Myrs for the full sample, 819 Myrs for the restricted sample) are similar --- suggesting that using the whole disk doesn't significantly bias the measurement of the interaction time --- the restricted sample gives significantly lower scatter in the estimated value of $t_0$, and a higher accuracy. The actual time since disk crossing is 844 Myrs, corresponding to the last disk-crossing of the dwarf. However, even though the simulation contains a relatively ``fast'' passage of the satellite, the perturbation is not a delta function, but rather occurs over some finite width of time, which depends on the orbital trajectory of the dwarf perturber and the geometry of a given particle in the disk with respect to the orbit of the perturber. We therefore expect there to be an inherent width to the true distribution of inferred interaction times.

In addition to different widths of interaction time, we also expect subtle differences in the time of peak perturbation for phase spirals in different regions of orbital space due to differences in stellar position in the disk at the time of interaction \citep{Gandhi_2022}. This can be seen in the right panel of Figure \ref{fig:times}, where we plot our measured values of $t_0$ as a function of $x_{\mathrm{act}}, y_{\mathrm{act}}$ with lighter regions corresponding to lower values of $t_0$ and darker regions corresponding to higher values of $t_0$. While there is large bin to bin variation, we do see coherent $x-y$ spiral structures in the estimated value of $t_0$ in the outer regions of the disk (larger $x_{\mathrm{act}}$--$y_{\mathrm{act}}$). 

In Figure~\ref{fig:fits} we zoom in on a small region in $J_\phi,\theta_\phi$ where the phase-spiral structure is the strongest in the simulations (this occurs around $J_\phi = 2300~\mathrm{kpc}~\mathrm{km}~\mathrm{s}^{-1}$, slightly outside of the solar neighborhood, but inside of the region of disk-crossing). In the main panel we show the background-subtracted phase spirals. It can be seen by eye how the characteristics of neighboring phase spirals (phase spirals in neighboring bins of $x_{\mathrm{act}},y_{\mathrm{act}}$) vary smoothly in strength, pitch angle, and morphology as a function of $J_\phi$ and $\theta_\phi$). In the bottom panel we show the ridgeline fits stacked in $J_\phi$. Here we can see that even for a fixed $J_\phi$ there are subtle shifts in the slope and pitch angle as a function of $\theta_{\phi}$. The same smooth change can be seen for fixed $\theta_\phi$ in the right panels, now as a function of $J_{\phi}$. 

The intuition for this variation is discussed more in Section \ref{sec:framework}, but briefly, stars in different $J_\phi$ and $\theta_\phi$ bins are on different orbits, and thus, will have been at different locations in the disk at the time of perturbation. An illustration of how this affects the measurement of $t_0$ is shown in Figure \ref{fig:traced_back}. Here, we plot the median interaction time for particles in bins of Cartesian coordinates $(x, y)$ in the plane of the disk (rather than $x_{\mathrm{act}}, y_{\mathrm{act}}$) at $t=-841$ Myrs. Thus, this plot shows bins of particles in real space at the disk crossing of the satellite galaxy, colored by the measured value of $t_0$ from the phase spiral at $t=0$. Here, we restrict to bins with $\sigma_t < 100$. As can be seen, at the time of disk crossing, there is a strong dipole pattern in the measured interaction times: stars that had their closest interaction longer ago show larger values of $t_0$ (redder), while stars that have yet to have their closest encounter with the satellite show shorter values of $t_0$ (bluer).\footnote{The disk rotation is counter-clockwise in this snapshot} This picture matches the intuition laid out in \cite{Gandhi_2022}: stars in different positions will experience slightly different bulk forces at slightly different times depending on their orientation with respect to the perturber. Local, vertical phase spirals will then wind at different speeds depending on the distribution of stellar frequencies. Both of these effects will lead to variations in phase spiral structure, even for a single, simple interaction. 

\begin{figure*}
\includegraphics[width=\textwidth,trim={18cm 10cm 19cm 8cm},clip]{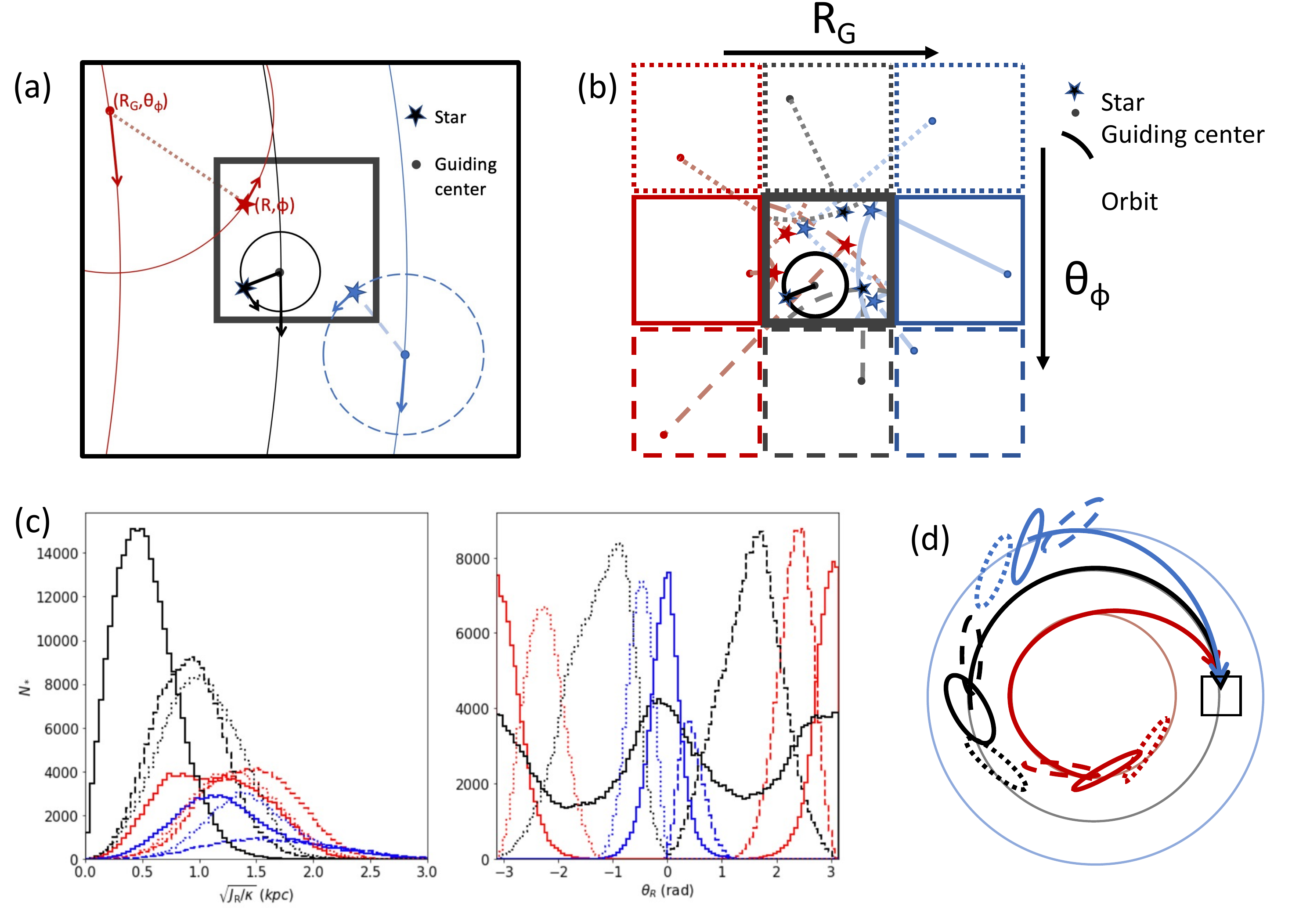}
\caption{Illustration of splitting a local sample in actions and angles, what it gains, and the biases introduced. \textbf{a)} Three stars that are currently in the observed sample (black box) at physical positions $(R,\Phi)$ in the disk (star), with their radial epicycles and guiding centres $(R_{\mathrm{G}},\theta_\phi)$ shown (dots). \textbf{b)} A set of stars projected into surrounding bins at larger (blue) or smaller (red) $R_{\mathrm{G}}$ and ahead (dotted bins) or behind (dashed bins) the local sample (black solid) in the direction of Galactic rotation. \textbf{c)} Example number counts in radial action (or eccentricity) and radial phase angle. The colors, and line styles show the distribution of $\sqrt{J_R/\kappa}$ and $\theta_R$ for stars in the nine different boxes in panel (b), which are all physically located within the black box, and projected across all nine. \textbf{d)} Illustration of where particles in a local sample (black box) might have been in the disk roughly half a solar rotation ago, coded by their $(R_G,\theta_\phi) $ bins from panels (b) and (c).} 
\label{fig:cartoon}
\end{figure*}

\section{Results II: Gaia DR3}

\subsection{Framework for Interpretation}
\label{sec:framework}

Before discussing the application of \texttt{ESCARGOT} to the \emph{Gaia} DR3 data, it is important to note some limitations of comparing our analyses of simulations to our analyses of the real data. For one, the simulations are split using particles from across the entire disk into bins of $J_\phi,\theta_\phi$, with each bin containing particles with the full range of eccentricities and at random places along their orbits between pericenter and apocenter (i.e. a random sampling in radial action and angle $(J_R,\theta_R$). When we make the equivalent projection of the real data --- where all stars are close to the Sun's position ---into $(J_\phi, \theta_\phi)$ bins, the same is not true. That is, by combining a positional selection (i.e. ``near the sun'') with a dynamical selection (i.e. in a bin of $J_\phi$), this naturally introduces selection effects that impact the distribution of observable radial actions and angles $J_R, \theta_R$ \citep[see, e.g.,][]{Hunt_2020}. 

Figure~\ref{fig:cartoon} provides a conceptual illustration to demonstrate how these selection effects impact the interpretation of dynamical quantities in the solar neighborhood. Panel (a) shows three stars (star markers) that are currently in the observed sample (black box) at positions $(R,\Phi)$ in the disk. Each is orbiting with its own pericenter and apocenter or, equivalently, different sizes of their epicyclic oscillations around their mean motion {\it guiding center} (filled circle marker) at $(R_G,\theta_\phi)$. The transformation to bins in $J_\phi,\theta_\phi$, is equivalent to binning the stars not at their observed positions, but rather at their guiding centers. 

Panel (b) of Figure \ref{fig:cartoon} shows how a set of stars project into the surrounding bins at larger (blue) or smaller (red) $R_G$ and ahead (dotted bins) or behind (dashed bins) the local sample in the direction of Galactic rotation. Visual inspection of these panels illustrates the selection effects imposed by the limits of the observing region on the populations in each of the bins. For example, stars in the solid red bin are limited to epicycles with radii between zero and twice the bin width, $\Delta R_G$. Moreover, they will only enter into the observed region towards the apocenters of their orbits.

Panel (c) of Figure \ref{fig:cartoon}, derived from a simulated sample, shows how the selection effects impose systematic, predictable biases in populations within each of the bins, in both radial action (or eccentricity) and radial phase. The colors, and line styles show the distribution of $J_R$ and $\theta_R$ for stars in the nine different boxes in panel (b).

Finally, it is interesting to think about the history of the bin populations. Panel (d) of Figure \ref{fig:cartoon} indicates where particles in the observed sample might have been in the disk roughly half a solar rotation ago, coded by their $(R_G,\theta_\phi) $ bins from panels (b) and (c). For each set of stars, the order imposed by their original guiding center phase is maintained, while they spread apart systematically around the disk according to their $R_G$ because of the associated trends in azimuthal time periods.  This effectively means that these different subset of stars, now all observed in the same volume, may have experienced, and be able to tell us about the perturbation from a different perspective.

In addition to the differences discussed above (i.e. related to selection effects), there is an added layer of idealization in the simulations we have presented. In these simulations, the particles are massless tracers of a static potential. Thus, they do not experience forces due to deformities induced in the dark matter halo from the passage of the dwarf galaxy \citep[found to be relevant in][]{Grand2022} nor do they experience self-gravity from the disk, or scattering from giant molecular clouds \citep{tremaine_23}. The effects of these simplifications are discussed further in Section~\ref{sec:discussion}. 

\begin{figure*}
\centering
\includegraphics[width=\textwidth]{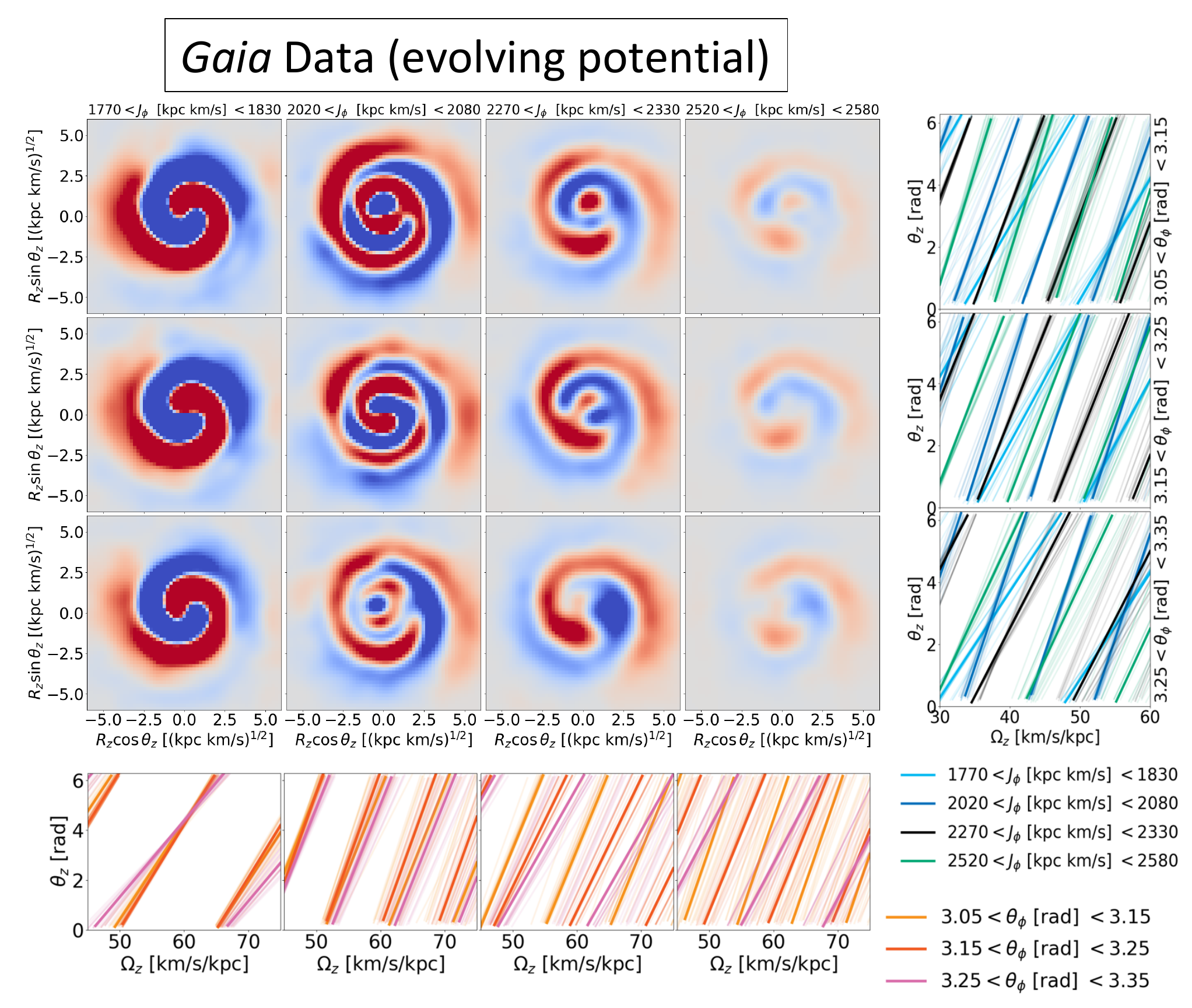}
\caption{\emph{Analysis of Gaia DR3 data --- center panel:} background subtracted phase spirals as a function of $J_\phi,-\theta_\phi$. \emph{frame:} Ridgeline fits to the spirals projected into $\theta_z = \Omega_z$ for different $\theta_\phi$ (right  panels) and $J_\phi$ (Bottom panels) bins. The lighter lines represent the fit to the data after bootstrap resampling. The center panel shows the variation in phase spiral morphology as a function of $J_\phi,-\theta_\phi$. In the bottom (left) panels the subtle shifts in phase spiral slope and pitch angle as a function of $\theta_\phi$ ($J_\phi$) can be seen.}
\label{fig:fits_gaia}
\end{figure*}

\begin{figure*}
\centering
\includegraphics[width=\textwidth]{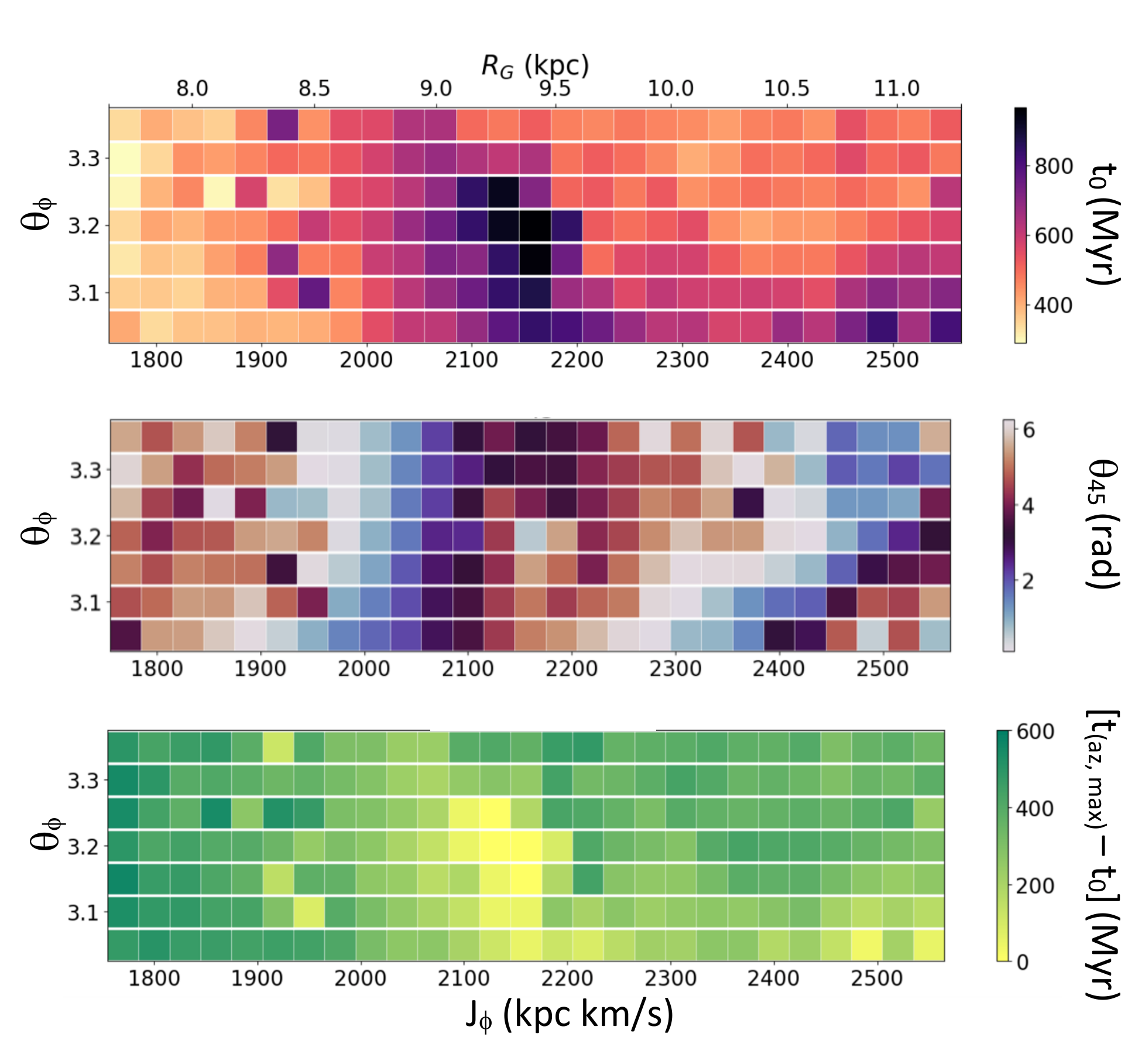}
\caption{\emph{Gaia} DR3 data \emph{top:} Measured interaction time ($t_0$) from boostrap resampling for each of the overlapping bins in $J_\phi-\theta_\phi$. The bins are labeled in $J_\phi$ (\emph{bottom}) and guiding radius (\emph{top}). Each bin span 60~\kpckms in $J_\phi$ and 0.01 rad in $\theta_\phi$.  The values of $t_0$ range from 288 to 966 Myrs with a median value of 524 Myrs. \emph{middle:} The fit value of $\theta_z$ at $\Omega_z = 45$, showing the variation in intercepts of the different fits for the same bins in $J_\phi-\theta_\phi$. \emph{bottom:} The offset between the time of maximum acceleration ($t_{a_z, \rm max}$) and the estimated time since interaction ($t_0$) for the same bins in $J_\phi-\theta_\phi$. The offsets range from -73 to 562 Myrs with a median value of 333 Myrs.}
\label{fig:fits_gaia_all}
\end{figure*}

\begin{figure}
\centering
\includegraphics[width=\columnwidth]{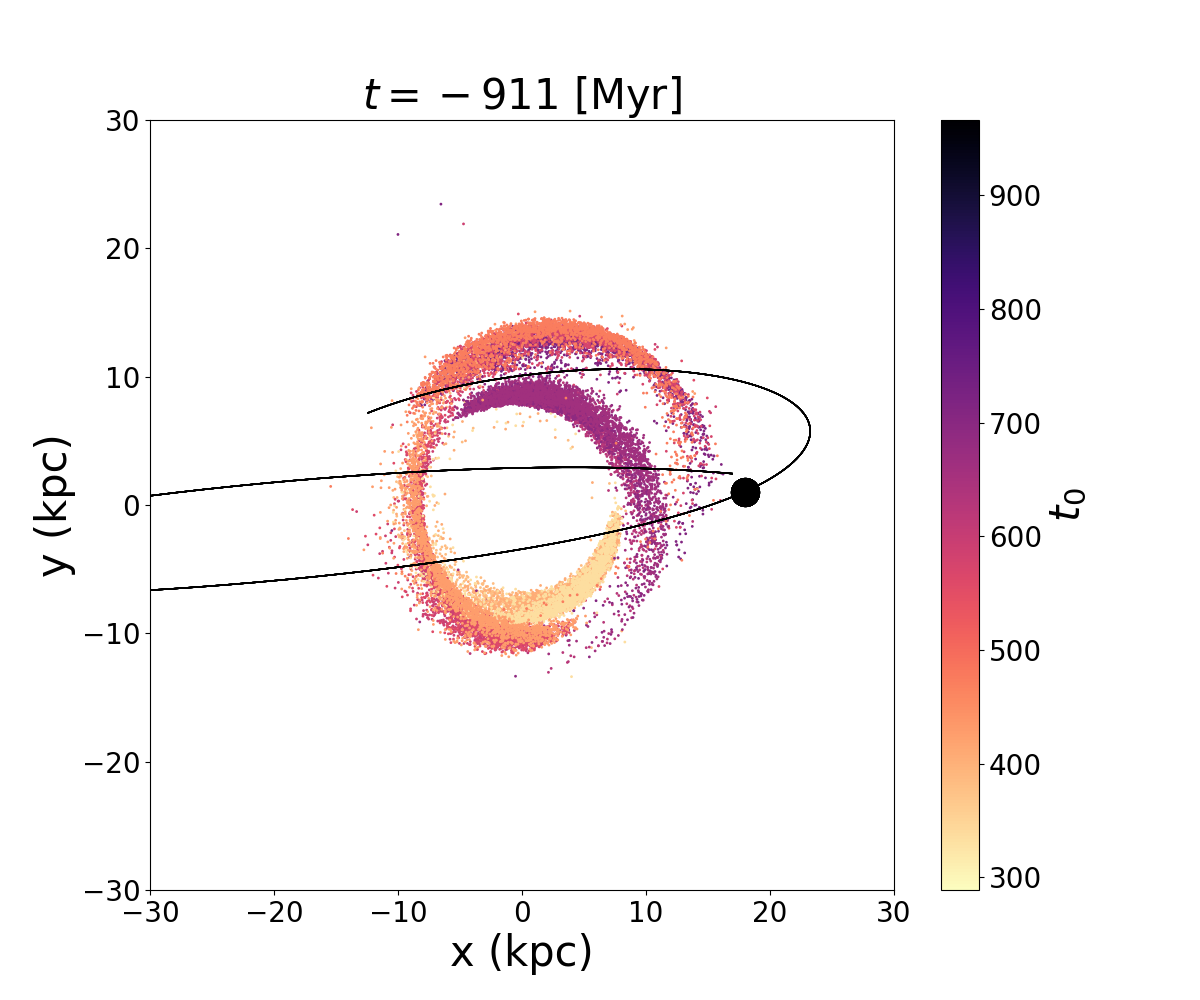}
\caption{\emph{Gaia DR3 data} in each of the $J_\phi - \theta_\phi$ bins in Figure \ref{fig:fits_gaia} backwards orbit integrated 911 Myrs. The data points are colored by the median $t_0$ (measured at $t = 0$ Myrs). The black line shows the orbit of Sagittarius and the black point shows its position at $t = -911$ Myrs.}
\label{fig:orbit_int}
\end{figure}

\begin{figure*}
\centering
\includegraphics[width=\textwidth]{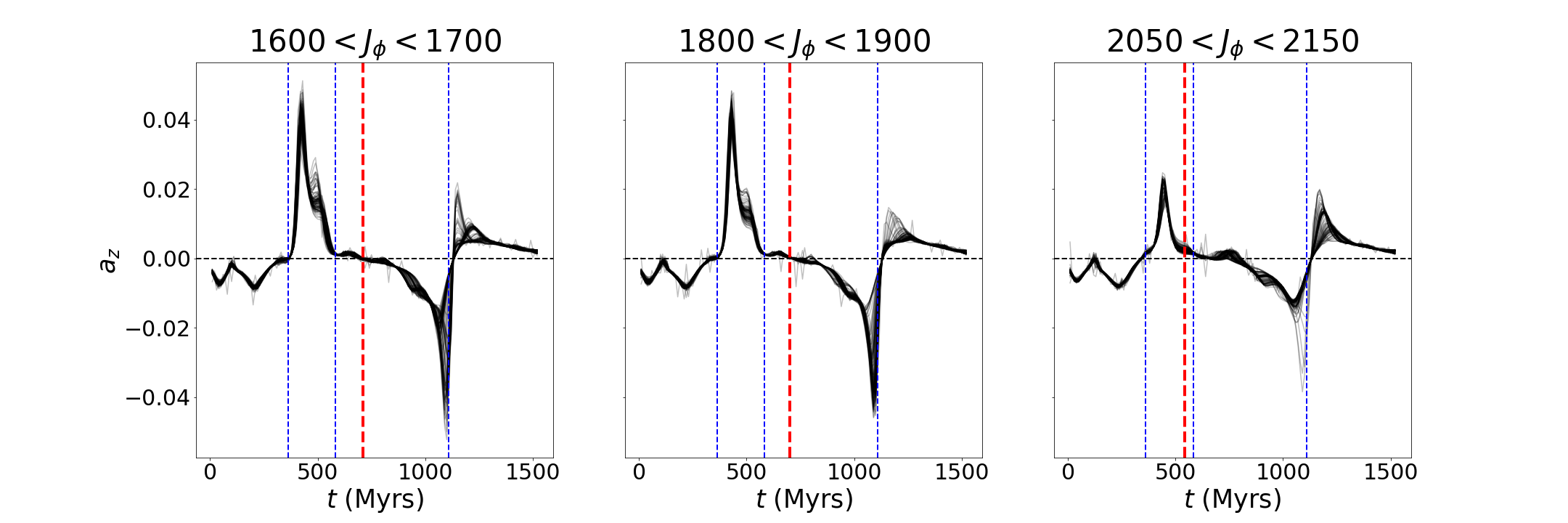}
\caption{Z-acceleration ($a_z$) from the satellite perturber vs. time for one hundred particles from the M1 simulation presented in \cite{Hunt_2021} in three bins of $J_\phi-\theta_\phi$ (black lines). The red dashed line marks the measured $t_0$ of each bin, while the blue dashed lines mark the disk crossings of the satellite at 362, 583, and 1110 Myrs}
\label{fig:z_acc_sim}
\end{figure*}
\subsection{Application of \texttt{ESCARGOT} to the Gaia DR3 Data}
\label{sec:gaia_data}

The {\it Gaia} DR3 data release has over 33 million sources with measured 6D parameters (including radial velocities). This represents a factor of four increase over the previous sample released as part of {\it Gaia} DR2. As shown in, \cite{Hunt_2022}, this higher fidelity sample has already revealed new features of the local phase spiral including the presence of previously undetected two-armed phase spirals in the inner Milky Way. Here, we demonstrate our ability to capture subtler changes in phase spiral structure using \texttt{ESCARGOT}. We bin stars into twenty-seven evenly-spaced bins in $J_\phi$ between $[1770, 2550]~\kpckms$ and seven evenly-spaced bins $\theta_\phi$ between $[3.05, 3.35]$ (where the sun is at $J_{\phi} = 2084~\kpckms$ assuming solar position $x_{\rm sun}=[-8.275, 0, 0.208]$ kpc and velocity $v_{\rm sun}=[8.4, 251.8, 8.4]$ km s$^{-1}$) (for an explanation, see Section~2 of \citealt{Hunt_2022}; source data from \citealt{Bennett:2019, Reid:2020, GRAVITY:2021}). The bins span $60~\kpckms$ in $J_\phi$ and $0.1$ rad in $\theta_\phi$, meaning that adjacent bins overlap in $J_\phi$ and $\theta_\phi$. We use a slightly smaller range in $R_z$ --- $2.25 < R_z~(\kpckms)^{1/2} < 4.25$ --- when applying our algorithm to the \emph{Gaia} data, which show more messiness in the inner regions of the phase spiral. This is most likely an artifact of calculating actions and angles in an approximate best-fit potential. For more discussion of the gravitational potential used see Appendix \ref{sec:appendix}.

We do not currently have enough resolution to robustly extract interaction times for the majority of the inner $m=2$ phase spirals, discovered in \cite{Hunt_2022}, which exist at lower values of $J_\phi$. Because of this, we leave detailed discussion of the interaction time of these modes to future work. However, we demonstrate \texttt{ESCARGOT}s ability to recover an interaction time for these $m=2$ for a single example bin in $J_\phi,\theta_\phi$ in Appendix \ref{sec:appendixB} where the phase spiral is well resolved. We find a very short interaction time, $t_0 = 133$ Myrs for this phase spiral, consistent with theories of a different physical origin for phase spirals in the inner disk \citep[see][]{Hunt_2022}. We plan to further investigate the interaction times and physical origins of the $m=2$ phase spirals in data from future, deeper, \emph{Gaia} data releases. 

In Figure~\ref{fig:fits_gaia}, we show the phase spirals and corresponding fits to twelve example slices in $J_{\phi}, \theta_{\phi}$, spanning the full range in $J_{\phi}, \theta_{\phi}$ explored. As can be seen by eye, the phase spirals vary in strength and pitch angle as a function of $J_{\phi}$ and $\theta_{\phi}$, with the central bin in $J_{\phi}$ -- $2020 < J_\phi < 2080$ (corresponding to the solar neighborhood) -- showing increased winding relative to the other bins. These variations are captured in the variations in slope and intercept seen in the corresponding fits on the right and bottom. The variation in pitch angle as a function of $J_{\phi}$ is most clearly captured in the different zero points of the various fits, while the variation in $t_0$ can be seen in the differences in the slope of the fit. The lighter lines are slopes from twenty bootstrap resamplings of the extracted ridgeline points and serve as a proxy for uncertainty on the slope. While some of the fits show significant scatter, there is still clear evolution between the slope of the fits in adjacent bins. 

The variation in the phase spirals as a function of $J_\phi,\theta_\phi$ can be seen more clearly in Figure~\ref{fig:fits_gaia_all}, which shows the estimated value of $t_0$ (top panel) and $\theta_{45}$ (middle panel) for each of the 189 bins.\footnote{$\theta_{45}$ is the value of $\theta_z$ from the fit at $\Omega_z = 45$} We recover a range of interaction times from $t_0 = 288$ to $t_0 = 966$ Myrs with a median of $t_0 = 524$ Myrs. However, the scatter in interaction times is not random, but rather shows correlated structure as a function of $J_\phi,\theta_\phi$. As with the simulated data (see right panel of Figure~\ref{fig:times}), we see a ridge-like pattern in the estimated values of $t_0$ (darker bins), which peaks around $J_\phi = 2050~\kpckms$, indicating that we can detect large-scale coherent patterns in phase spiral structure across $J_{\phi}$ \citep[as also shown in][]{frankel_22,antoja_22}. It is striking that we see systematic variation with both $J_\phi$ and $\theta_\phi$ across the data. The pattern is coherent across the Galactic plane and mirrored in both $t_0$ and $\theta_{45}$ and reminiscent of the large-scale patterns in stellar velocity discovered  with \emph{Gaia} DR2 \citep{Katz2018}. This suggests that these large-scale variations in spiral morphology could be intimately linked to variations in average $v_R$, or bulk kinematics of the disk.

\subsection{Investigating Sagittarius as a possible culprit}

\subsubsection{Backward Integration}

One possible origin for the phase spirals in the disk is an interaction with the Sagittarius dwarf galaxy, which has been shown to be the dominant external perturber in this region of the disk \citep{Laporte_2019,Banik_2022}. 
In order to investigate whether the range in the measured values of $t_0$ in the different $J_\phi$ bins could be attributed to differences in disk location at the time of Sagittarius' recent disk crossings (see Section \ref{sec:framework}), we use backwards orbit integration of Sagittarius \citep[with its present location taken from][]{bennett_22}, along with a random selection of stars from each $J_\phi,\theta_\phi$ bin in Figure~\ref{fig:fits_gaia}. The orbit integration is done using the package \texttt{gala} \citep{gala} in the best fit potential found in Section~\ref{sec:appendix}.  We find a disk crossing time of $t_{\rm cross} \sim 911$ Myr ago (there is another crossing at 338 Myrs, but this occurs much further out in the disk). Given the uncertainties in the present-day position of Sagittarius, the error bars on this estimate are on the order of $150$ Myrs (but note this does not include uncertainties in the gravitational potential of the Milky Way; see discussion below). However, even with the error bars, this estimate is longer than the majority of our measured values of $t_0$ which range from $288$ to $966$ Myrs with a median of $524$ Myrs

The positions of stars at $t_{\rm cross}=-911$ Myrs relative to Sagittarius (\emph{black point}) can be seen in Figure \ref{fig:orbit_int}, where the particles are colored by the measured interaction time of a given bin (analogous to Figure \ref{fig:traced_back}). For visual clarity, we only plot stars from the twelve bins shown in Figure \ref{fig:fits_gaia}, which span the range of $J_\phi,\theta_\phi$ bins we fit. We find that stars from bins with larger values of $t_0$ were generally closer to Sagittarius at the time of disk crossing in-line with our intuition that stars which interacted earlier should have longer measured times since interaction. 

To investigate the apparent discrepancy between $t_0$ estimated from the phase spirals and $t_{\rm cross}$ from backwards integration we look at whether the interaction between stars in different $J_\phi,\theta_\phi$ bins could have occurred at different times. We track the average time of maximum acceleration from Sagittarius, $t_{a_z, \rm max}$, for particles in each bin during the backwards integration. The bottom panel of Figure~\ref{fig:fits_gaia_all}, shows the difference between $t_0$ and $t_{a_z, \rm max}$, which we expect to roughly correspond to the interaction time between Sagittarius and the stars in a given bin. As can be seen in Figure~\ref{fig:fits_gaia_all}, most bins show a significant positive offset, indicating that, even in this case, our estimated interaction times are shorter than expected from orbit integration. We find a wide range in offsets from 0 Myrs to greater than 550 Myrs, with four bins showing negative offsets (-73 to 0 Myrs). In addition, these offsets show correlated structure, with the smaller offsets occurring around $J_\phi = 2150$. This corresponds to the same ridge-like region in $t_0$ in the top panel, where we also see tighter wrapping in the phase spirals (see Figure \ref{fig:fits_gaia}). Overall we measure offsets between -73 to 562 Myrs with a median value of 333 Myrs. However, it should be noted that these estimates have uncertainties on the order of 250 Myrs from uncertainties in the orbit of Sagittarius alone. There are additional uncertainties on the shape of the potential or time-dependent effects from the impact of the LMC \citep{Vasiliev_2020}. This makes it hard to determine whether the origin of this offset suggests a dynamical mechanism, either due to Sagittarius or another driver of phase spiral formation, or is attributable to our combined uncertainties on the orbital history of Sagittarius.

\subsubsection{Comparison to fully self-consistent simulations}

It is tempting to interpret the offset between our estimated ($t_0$) and expected ($t_{a_z,\rm max}$) interaction times as ruling out Sagittarius as a possible culprit for inciting the phase spirals. 
In order to test the validity of this conclusion, we repeat our measurements using the fully self-consistent simulations introduced in \cite{Hunt_2021} (see within for details of the simulation set-up). We apply our algorithm to phase spirals at snapshot 702 -- the same snapshot defined as the ``present-day'' in \cite{Hunt_2021}, occurring at 6.87 Gyr. We chose this snapshot due to the presence of clear one-armed spirals in the simulations. The orbit of the satellite perturber in these simulations is not constrained to match our current understanding of the specific orbit of Sagittarius in the Milky Way \citep[e.g.][]{Vasiliev_2020}. At the time of this snapshot, the satellite has undergone three recent disk passages, one at approximately 1110 Myrs ago and two more recent crossings at 362 and 583 Myrs ago.

We reconstruct the host galaxy potential and calculate actions using \texttt{agama } \cite{agama}. We use multipole expansions for the bulge and dark matter halo, and a Cylindrical spline for the disk. Then we calculate actions and angles from the simulation using \texttt{agama}'s \texttt{ActionFinder}. However, we `flatten' the vertical distribution such that the vertical actions and angles are with respect to the local midplane, by subtracting off the median $z$ and $v_z$ across the disc. Assuming a global midplane instead of taking into account local corrugations leads to systematic biases in the derived actions and angles \citep{beane_19}. 

We select stars in a narrow spatial region to better mimic the data (see Section \ref{sec:gaia_data}). We select stars between $-8.678 < x$ (kpc) $< -7.678$ and  $|y| < 1$ (kpc). Because of this narrow selection and the resolution limits of the simulations, we are only able to resolve phase spirals in a limited number of $J_\phi$ bins. We measure phase spirals in three bins of 100~\kpckms in $J_\phi$ centered at $[1650, 1850, 2100]~\kpckms$ at $\theta_\phi = 3.2 \pm 0.15$ rad. We extract the ridgeline between $1.0 < R_z~(\kpckms)^{1/2} < 4.25$ due to low resolution in the outer phase spiral regions. The extracted values of $t_0$ compared with the vertical acceleration due to the satellite, $a_z$, for particles in each bin can be seen in Figure~\ref{fig:z_acc_sim}. Each line corresponds to a different particle in a given $J_\phi,\theta_\phi$ bin. The vertical width of the various spikes indicates the amount of variation in the strength of $a_z$, even for particles in the same $J_\phi,\theta_\phi$ bin. 

As can be seen, the extracted values of $t_0$ lie somewhere in between the more recent and the earlier passage of Sagittarius with an offset between 400--600 Myrs from the earliest passage, similar to the offsets we measure in \emph{Gaia} (Section \ref{sec:gaia_data}). Rather than corresponding to the peak of $a_z$, the measured value of $t_0$ appears to correspond to the end of the interaction ($a_z$ = 0). This may explain why we do not measure an interaction time consistent with the more recent spike in $a_z$, which occurs between the two more recent disk crossings, as this interaction only recently ended and therefore may not have had enough time to develop a well-formed spiral. We do see a dipole in the phase spiral before performing the midplane subtraction, which could be the signature of the beginning of the formation of a new phase spiral caused by the more recent interaction.  

This suggests that the offsets between $t_0$ and the time of maximum $z-$acceleration $(t_{a_z, \rm max})$ we find in the \emph{Gaia} data may have physical origins. Since we do not see equivalent offsets in the tracer particle simulations (Section~\ref{sec:intro}), we believe these offsets are due to more complicated interactions between the satellite, Milky Way dark matter halo, and self-gravity in the disk itself, which are not captured in the simplified simulations (Section~\ref{sec:framework}). Note that the simulation of \cite{Hunt_2021} does not contain giant molecular clouds or dark matter subhalos, and thus the discrepancy between measured age and impact time (in this model) cannot be explained by such scattering, as suggested in \cite{tremaine_23}. While a full analysis of this discrepancy is beyond the scope of this work, we flesh out current theories and plans for next steps in the following sections. 

\section{Summary and Discussion}
\label{sec:discussion} 

\subsection{Summary}

In this work, we describe \texttt{ESCARGOT}, an algorithm designed to characterize phase spirals in action--angle space with the goal of using the spirals themselves to deepen our understanding of the dynamical mechanisms that cause them. \texttt{ESCARGOT} is designed to extract global quantities from a given spiral including the spiral mode ($m$) and the time since formation ($t_0$). We apply \texttt{ESCARGOT} to a simple simulation consisting of test particles in a static potential that undergos a single interaction with a dwarf galaxy perturber on a rapid orbit. We show that \texttt{ESCARGOT} does a good job of characterizing the phase spirals and recovering the interaction time of the satellite from the median measured interaction time across a range of phase spirals. However, we also show that there exists scatter in individual estimates that show correlated coherent patterns across a range of $J_{\phi}$ and $\theta_{\phi}$ even in the simple example of a single perturbation.

Next, we apply \texttt{ESCARGOT} to observed data from the recent \emph{Gaia} DR3 data release. Binning in $J_\phi$ and $\theta_\phi$ we find a range of interaction times that also vary coherently with $J_{\phi}$ and $\theta_{\phi}$, similar to the simulation. However, unlike the simulation, the recovered times are significantly offset from the time of last disk-crossing of Sagittarius estimated using backwards orbit integration of the remnant. 

Repeating this analysis in the set of more realistic simulations described in \cite{Hunt_2021}, we find a similar offset between the measured interaction time ($t_0$) and the periods of maximum $z$-acceleration induced by the simulated satellite. This suggests the offset could be due to the physical effects of dark matter halo wakes \citep{Grand2022} or the influence of self-gravity in the disc. The $N$-body simulation does not contain giant molecular clouds or dark matter subhalos, and thus they cannot be invoked to explain the offset in this model, although their influence may be important in the Milky Way \citep{tremaine_23}. We defer a more thorough exploration of this to further work.

\subsection{Comparison with Contemporaneous Work}

The analysis presented here shares conceptual similarities with the recent works by \cite{frankel_22} and \cite{antoja_22}, which both present measurements of $t_0$ from \emph{Gaia} data at a range of locations in the disk, as well as a slightly earlier paper by \cite{Widmark_2022}, which presents similar analysis. As discussed in Section \ref{sec:gaia_data}, our estimated values of $t_0$ range from $288$ to $966$ Myrs with a median of $524$ Myrs, in good agreement with these works as well as the original analysis by \cite{Antoja_2018}, which put the time since interaction between 300 and 900 Myrs with $t_0 = 500$ Myrs. 

\cite{frankel_22} measures the interaction times of phase spirals in $\sqrt{J_z}, \theta_z$ space for stars in \emph{Gaia} eDR3 using a parameterized model of the density of stars in $\theta_z-\Omega_z$. They find a similar wave-like pattern in the measured interaction times, although the peak is shifted to lower $J_\phi$. While their algorithm is not capable of modeling the $m=2$ modes, they find indications of such modes in the model residuals at low $J_\phi$ as shown in \cite{Hunt_2021} and Appendix \ref{sec:appendix}. They also find relatively short interaction times across all the bins, ranging from $t_0 \in [200, 600]$ Myrs. 

\cite{antoja_22} measures the interaction times in $z-v_z$ rather than $\sqrt{J_z}, \theta_z$ space for stars in \emph{Gaia} DR3 using an edge detection algorithm. They find a similar range of interaction times, $t_0 \in [300, 900]$ Myrs, which vary coherently as a function of $J_\phi$ (although they do not see evidence for the inner $m=2$ modes). 

In a slightly earlier paper --- which also modeled the phase-space density of the spiral --- using \emph{Gaia} eDR3 data, \cite{Widmark_2022} found $t_0 \in [350, 767]$ Myr. While the distribution of times is plotted as a function of physical position rather than angular momentum and azimuthal phase angle, patterns of large-scale coherence between the estimated interaction times are visibly present. 

The broad agreement between these different algorithms strengthens the claim that the local phase spiral has a relatively short interaction time $t_0 < 1$ Gyr and shows coherent large-scale variation as a function of both physical location, $J_\phi$, and $\theta_{\phi}$. This picture agrees well with the intuition that emerges from our analysis of high-resolution simulations, which also generically shows large-scale coherence in the measured values of $t_0$, and in the fully self-consistent simulations presented in \cite{Hunt_2021}, shorter than expected interaction times.  

However, it should be noted, that other efforts place the perturbation further in the past. These works differ from those quoted above, as they estimate $t_0$ using analysis methods that do not focus on direct unwinding of the phase spiral. \cite{Darling_2018} find a interaction time of around 1 Gyr by comparing the present-day shape of the \emph{Gaia} phase spiral to a set of self-consistent simulations. This estimate is similar to the estimate found by \cite{Bland_Hawthorn_2021}, which also bases their estimate on comparison to a set of tailored hydrodynamical N-body simulations of the Milky Way--Sagittarius merger, and by \cite{Ruiz-Lara_2020}, who estimate the interaction time from looking at bursts of star formation. Lastly, \cite{Li_2020} puts a constraint on $t_0 > 500$ Myrs by looking at phase-mixing in arches.

\subsection{Phase Spiral Formation Mechanisms} 

\subsubsection{It could still be Sagittarius}

Within the picture where Sagittarius is responsible for driving the observed phase spirals, there are many possible causes for the lag between the measured values of $t_0$ for the phase spirals and the passage of Sagittarius. The first \citep[recently discussed in][]{Grand2022} is that the phase spirals are being excited by large-scale wakes in the dark matter halo rather than the direct torque from the satellite passage. These wakes could have intricate large-scale structure that forms dynamically on different time scales, leading to complicated interaction pattern across the disk.  

Another possibility is that in a realistic disk, self-gravity -- rather than the direct torque induced by the satellite passage -- is the most significant driver of phase spiral formation and evolution. In the test particle model described above, the perturbation to the disc does not change the underlying potential. When stars are pulled away from the mid-plane of the Galaxy they oscillate around the previous and fixed center of the vertical potential. In a more realistic simulation such as the $N$-body model from \cite{Hunt_2021}, or the Milky Way itself, when stars are pulled away from the previous midplane, the center of the vertical potential will move (slightly) with them, as the gravitational force from the perturbed stars themselves contribute to the disc potential. This will change the local restoring force, slowing the onset and subsequent winding of the phase spirals. This effect would also fade as perturbed populations mix radially and azimuthally.

In addition, both the impact of realistic self-gravity and large-scale dark matter wakes could drive vertical wave-like patterns in the disk which re-excite local phase spirals at different times and on shorter time scales. Indeed, signatures of oscillations in the midplane of the Galaxy are abundant in recent data sets \citep{Newberg2002, Widrow:2012, Williams2013, Carlin2013, PW2015, Xu2015, Antoja_2018, antoja_22}. 

Finally, the discrepancy could simply be caused by uncertainties in the measurements of the current phase-space positions of Sagittarius as well as inaccuracies in our potential model for the Milky Way. Indeed, if we can understand how such uncertainties would manifest in our distributions of $t_0$ we could use that knowledge to refine both.


\subsubsection{Other origins}

These mechanisms differ somewhat from the mechanism recently proposed by \cite{tremaine_23}, which attributes the phase spirals to a Gaussian process consisting of many weak perturbations to the distribution function due to giant molecular clouds. While this mechanism is good at generating phase spirals with short interaction times $t_0 < 500$ Myrs, it remains unclear whether they can generate the large-scale coherent structure seen in the distribution of measured interaction times in both the \emph{Gaia} data (Figure \ref{fig:fits_gaia_all}) and the idealized simulations (Figure \ref{fig:times}), which does not contain small-scale structure in the disk.

\section{Conclusions }
\label{sec:conclusion}

This paper centers around four key points:
\begin{enumerate}
\item We present \texttt{ESCARGOT}, an algorithm capable of characterizing the nature of phase spirals seen in local stellar data in terms of three parameters: $m$, $t_0$ and $\theta_{45}$ corresponding to the spiral mode, the time since interaction, and the intercept of the best-fit-line to the spiral.
\item We apply \texttt{ESCARGOT} to the {\it Gaia} data in $27 \times 7$ bins in $J_\phi, \theta_\phi$. We find ranges of $t_0 \in [288, 966]$ Myr and $\theta_{45} \in [0, 2\pi]$ rad for our parameter estimates.
\item We find that these estimates vary systematically and coherently across the range of $J_\phi, \theta_\phi$ bins analyzed.
\item These coherent and correlated results lead us to conclude that: (i) the phase spirals contain meaningful information about one or several past interactions; and (ii) that \texttt{ESCARGOT} is successfully extracting this information.
\end{enumerate}

Our preliminary investigation of the possibility that the spirals are excited by the Sagittarius dwarf led to consistent, but as yet inconclusive, results. Nevertheless, the promise of using phase spirals as probes of the Milky Way's past and mass distribution cannot be overstated. 

In practice, the Milky Way phase spirals may originate from a variety of effects combined.  It is not yet clear if only one influence is dominant, or whether we shall have to disentangle the origin of specific phase spirals from a complex combination of perturbative forces. Regardless, quantitative measurements of phase spiral properties, such as presented in this work \citep[see also][]{frankel_22,antoja_22} will be vital in future attempts to exploit the rich information contained within these striking dynamical phenomena.

\section*{Acknowledgements}
We make use of data from the European Space Agency (ESA)
mission $Gaia$ (\url{http://www.cosmos.esa.int/gaia}), processed
by the $Gaia$ Data Processing and Analysis Consortium (DPAC,
\url{http://www.cosmos.esa.int/web/gaia/dpac/consortium}). Funding for the DPAC has been provided by national institutions, in particular the institutions participating in the $Gaia$ Multilateral
Agreement. This research made use of \texttt{astropy}, a community-developed core Python package for Astronomy \citep{astropy-1,astropy-2}, and the galactic dynamics Python packages \texttt{agama} \citep{agama}, \texttt{Gala} \citep{gala} and \texttt{galpy} \citep{bovy_15}.

EDF received support from the U.S. Department of Energy under contract number DE-AC02-76SF00515 to SLAC National Accelerator Laboratory.

\bibliography{sample631}{}
\bibliographystyle{aasjournal}

\begin{figure*}
\centering
\includegraphics[width=\textwidth]{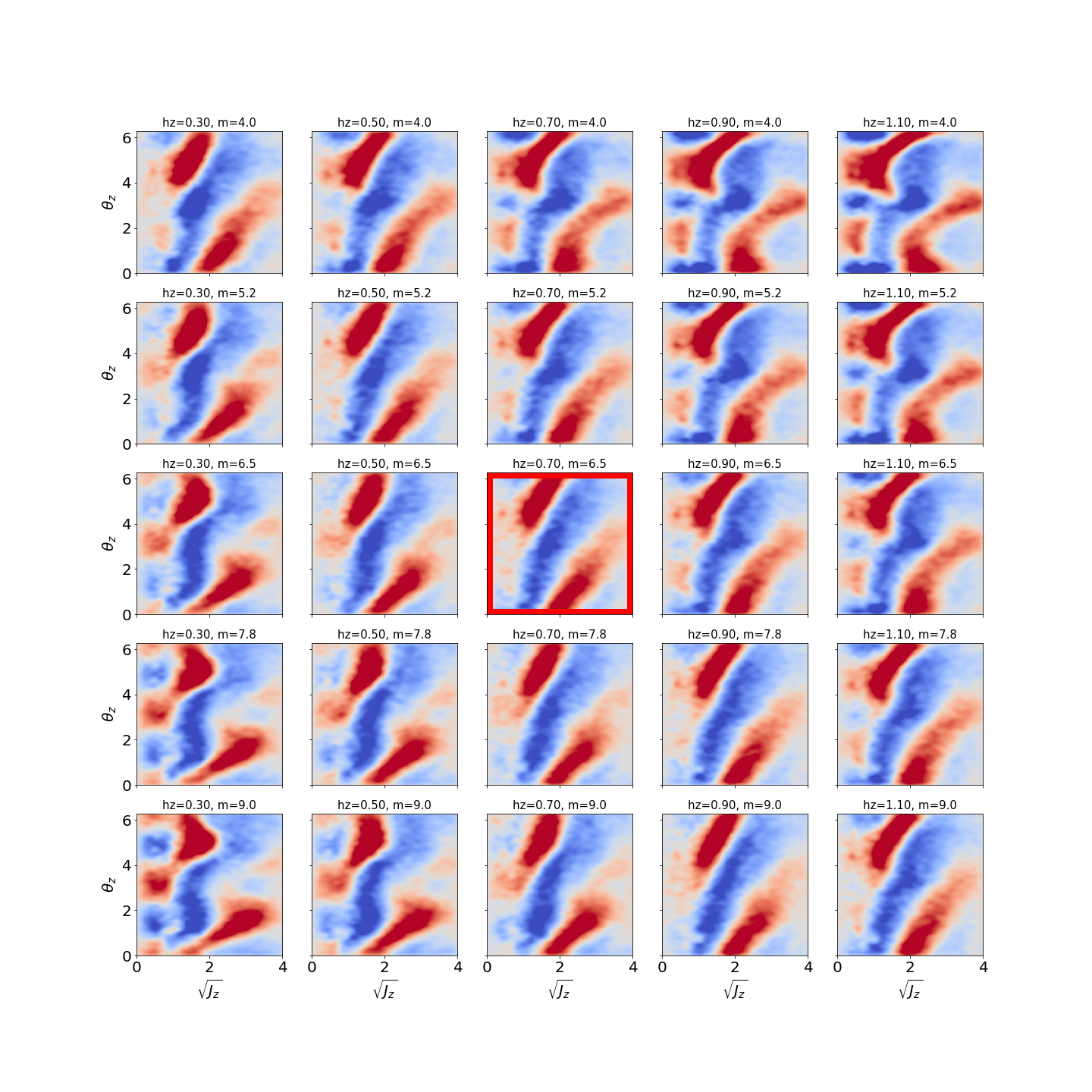}
\caption{\emph{Gaia} data showing phase spirals unwrapped in vertical action--angle coordinates $(J_z, \theta_z)$ as computed in a 2D grid of mass models for the Milky Way. 
Each panel represents the phase spiral as computed in a potential model with the parameters indicated in each panel title, where $h_z$ is the disk scale height in kpc   and $m$ is the disk mass in $10^{10}~\Msun$;
The rows represent increasing scale height and the columns represent increasing disk mass.}
\label{fig:fits_potential}
\end{figure*}

\appendix 
\section{Determining Milky Way Potential}

\label{sec:appendix}

In order to run \texttt{ESCARGOT} on the {\it Gaia} data, we first have to decide on a mass model for the Milky Way to calculate the actions and angles. 
In order to find a suitable potential, we run the action calculations in a grid of potential parameter values where we vary the disk scale height ($h_z$) and the disk mass ($m$) at fixed circular velocity at the solar circle $v_{c, \odot} = 229~\kms$ \citep[][we adjust the halo mass to keep this fixed]{Eilers:2019}.
Figure~\ref{fig:fits_potential} shows an example of a spiral unwrapped and plotted in the space of vertical action $J_z$ and angle $\theta_z$ for this 2D grid of Milky Way potential models.
From this grid, we picked a potential model (visually) in which the spiral is close to a straight line in the space of $\sqrt{J_z}, \theta_z$ (the panel outlined with red line) and adopt this as our \textit{Fiducial potential} used in Section~\ref{sec:gaia_data}.
In bad models for the potential, the spiral shows ``wiggles''  and departures from a linear relation in this space (e.g., top right panel of Figure~\ref{fig:fits_potential}).

\begin{figure*}
\centering
\includegraphics[width=\textwidth]{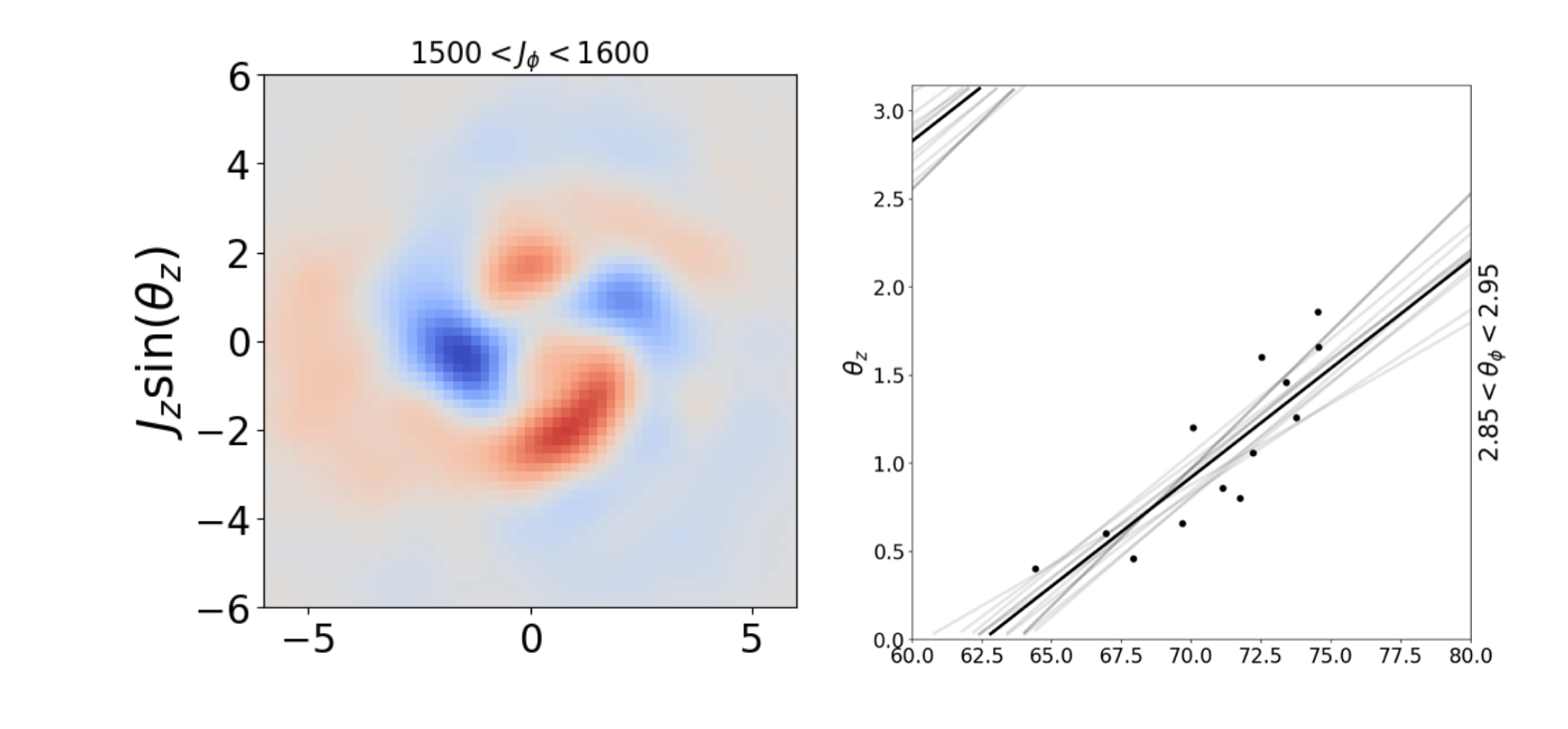}
\caption{Example $m=2$ phase spiral \emph{(left)} and corresponding fit to extracted ridgeline \emph{(right)}. For this spiral we get an interaction time of $t_0 = 133$ Myrs}
\label{fig:n_2}
\end{figure*}

\section{m=2 phase spiral Modes}

\label{sec:appendixB}
An example of a single fit of \texttt{ESCARGOT} to an $m=2$ phase spiral can be seen in Figure \ref{fig:n_2} for a bin centered at $J_\phi = 1550~\kpckms$ and $\theta_\phi = 2.8$. We recover a very short interaction time of $133$ Myrs, consistent with the theory proposed in \cite{Hunt_2022}, that the $m=2$ modes are potentially being excited by a different mechanism than the $m=1$ modes (i.e. perturbations from spiral arms or the galactic bar), which would occur over much shorter time scales. We leave a fuller investigation of the $m=2$ modes and their origins to future work. 



\end{document}